# Experimental Evidences for Static Charge Density Waves in Iron Oxy-pnictides

A. Martinelli[1], P. Manfrinetti[1,2], A. Provino[1,2], A. Genovese[3], F. Caglieris,[1,4], G. Lamura[1], C. Ritter[5], M. Putti[1,4]

[1] *SPIN-CNR, C.so Perrone 24, I- 16152 Genova – Italy*

[2] *Department of Chemistry and Industrial Chemistry, Università di Genova, Via Dodecaneso 31, I-16146 Genova, Italy*

[3] *Biological and Environmental Sciences and Engineering Division, King Abdullah University of Science and Technology, Thuwal 23955-6900 Jeddah, Saudi Arabia*

[4] *Department of Physics, Università di Genova, Via Dodecaneso 33, I-16146 Genova, Italy*

[5] *Institute Laue - Langevin, 6 rue Jules Horowitz, 38042 Grenoble Cedex 9 – France*

**Abstract**

In this Letter we report high-resolution synchrotron X-ray powder diffraction and transmission electron microscope analysis of Mn-substituted LaFeAsO samples, demonstrating that a static incommensurate modulated structure develops across the low-temperature orthorhombic phase, whose modulation wave-vector depends on the Mn content. The incommensurate structural distortion is likely originating from a charge-density-wave instability, a periodic modulation of the density of conduction electrons associated with a modulation of the atomic positions. Our results add a new component in the physics of Fe-based superconductors, indicating that the density wave ordering is charge-driven.

**Text**

Since the discovery of superconductivity in Fe-based tetrahedral layered systems, there have been much efforts to understand the tangled interplay among structural, electronic and magnetic properties dominating these materials. In particular, the origin of the tetragonal to orthorhombic structural transition in Fe-based superconductor materials is still debated. Indeed, the lattice distortion is a secondary response, driven by collective electronic degrees of freedom (spin, charge, orbital); which of them plays a dominant role is controversial. Moreover, the structural transition is preceded by a nematic order, a state where the electronic properties display a 2-fold rotational symmetry about the tetragonal *c*-axis [1].

The nesting between hole and electron Fermi surface pockets is one of the most prominent phenomena characterizing the physics of the Fe-based superconductors and their parent compounds [2,3,4]. In principle, Fermi surface nesting can destabilize the system, leading to a charge density wave (CDW) or to a spin density wave (SDW). In the first case the structure can spontaneously distort on account of the electron-ion Coulomb interaction and the CDW can give rise to a structural transformation, resulting in many cases in an incommensurate structure, since the lattice periodicity is related to the Fermi wave vector. Conversely, the SDW depends on electron-electron interactions and in this case the system is not necessarily prone to a structural transformation.

In chronological sequence, specific heat measurements first indicated a phase transition at around 150 K in LaFeAsO; at the same time, the earliest optical measurements detected no new phonon modes or their splitting throughout this transition and the authors concluded that the phase transition did not involve a structural distortion [4]. On these bases, it was speculated that a SDW transition takes place, ruling out a structural transition and the occurrence of CDW order as well. Later, these conjectures were at least partly contradicted by neutron powder diffraction analyses [5], that definitively ascertained the occurrence of a structural transformation in pure LaFeAsO located at $T_s \sim 155$ K. On the other hand this structural transition was found to be combined with an antiferromagnetic ordering characterized by an in-plane tetragonal magnetic wave-vector $\mathbf{k}_{magnetic} = (\pi,\pi)$ and a reduced ordered magnetic moment at the Fe site ($< 1$ $\mu_B$), perfectly consistent with the predicted SDW state. Probably for these reasons, the SDW scenario prevailed, even though the occurrence of a structural transition would strengthen the possibility for a CDW state. Remarkably, a modulated superstructure usually develops when a CDW is involved in a structural transformation but, so far, no incommensurability had been detected in the low-temperature orthorhombic phase of the 122- and 1111-type compounds. Probably for this reason the CDW scenario has been almost completely neglected. Indeed, theoretical investigations [6,7] and symmetry mode analysis [8] point to an electronic origin of this tetragonal-to-orthorhombic transformation. Furthermore subsequent analyses revealed the occurrence of a nematic order pre-empting and possibly triggering the structural transformation; nematic order consists of the spontaneous breaking of the electronic symmetry between the *x* and *y* directions in the Fe-plane, but not of the underlying (tetragonal) lattice. As a consequence, several physical properties (transport, magnetic and optical properties) display a lower symmetry in the nematic state than the one characterizing the hosting crystal lattice; furthermore, nematic fluctuations could also be relevant to the pairing mechanism [9]. Despite very numerous investigations, the exact origin of the nematic state and of the tetragonal-to-orthorhombic structural transformation in 122- and 1111-type Fe-based superconductors still represents an enigma and different theoretical models were developed calling into question charge, orbital and spin degrees of freedom [1].

Within this class of materials, the optimally electron-doped La(Fe$_{1-x}$Mn$_x$)As(O$_{0.89}$F$_{0.11}$) system stands out for its peculiar behaviour; in fact, these compounds exhibit an anomalously huge and unexplained suppression of superconductivity by Mn-substitution as small as $x$ ~0.2%, whereas static magnetism is recovered for $x$ ~0.1% [10,11]. Remarkably, in (La$_{2-x}$Ba$_x$)CuO$_4$ (and some related systems as well) magnetism sets in and superconductivity is anomalously suppressed in a similar way for a hole concentration $x$ ~ ⅛. More interestingly, in these Cu-based systems a static CDW takes place, that is associated with the suppression of superconductivity and stems from the pinning of dynamical charge density waves [12,13,14].

For this reason, we focused our attention on the Mn-substituted LaFeAsO system in order to investigate the possibility that the physics of Fe-based superconductors could involve also the CDW state, among all the other components.

Polycrystalline samples were prepared by a solid state reaction at high temperature, sealing stoichiometric amounts of powders (LaAs, Fe$_2$O$_3$, MnO, Fe) in evacuated silica tubes.

Synchrotron X-ray powder diffraction analysis and high resolution transmission electron microscopy (HRTEM) were applied to investigate the structural properties of La(Fe$_{1-x}$Mn$_x$)AsO samples ($x$ = 0.02 and 0.04) between 290 K and 10 K. X-ray scattering data ($\lambda$ = 0.400 Å) were collected at the ID22 high-resolution powder diffraction beamline of the European Synchrotron Radiation Facility (ESRF) in Grenoble, France. The crystal structure was refined using the program FULLPROF [15]. Samples for transmission electron microscope observation (TEM) were prepared by mechanical milling with subsequent deposition of crystallites onto ultra-thin carbon coated TEM grids. HRTEM observation was performed by a FEI Titan Cube microscope, working at 300 kV, equipped with a Schottky electron source, a CEOS spherical aberration corrector of the objective lens, which allows to reach a sub-angstrom resolution (0.9 Å), and a post column Gatan Image Filter (GIF) Tridiem. HRTEM analysis was carried out down to 94 K by using a GATAN double tilt cryogenic-holder. DC magnetic measurements were achieved using a Quantum Design MPMS2 dc-SQUID magnetometer, revealing that the inflection points of the magnetization curves occur approximately around ~ 105 K and ~ 35 K for $x$ = 0.02 and 0.04, respectively; preliminary high-intensity medium-resolution neutron powder diffraction data collected on the sample La(Fe$_{0.98}$Mn$_{0.02}$)AsO fix the SDW transition temperature $T_{SDW}$ = 99.5±0.5 K (see the Supplementary Information for details).

The sample La(Fe$_{0.98}$Mn$_{0.02}$)AsO reveals a spectacular suppression of the structural transformation, down to $T_s$ ~ 115 K (for comparison in pure LaFeAsO $T_s$ ~ 155 K [5]). All the reflections of the diffraction pattern collected at 10 K are fitted by Rietveld refinement using an orthorhombic *Cmme* structural model; nonetheless, two extremely weak Bragg reflections at Q ~ 2.04 Å$^{-1}$ and 4.08 Å$^{-1}$ are not accounted for (Figure 1).

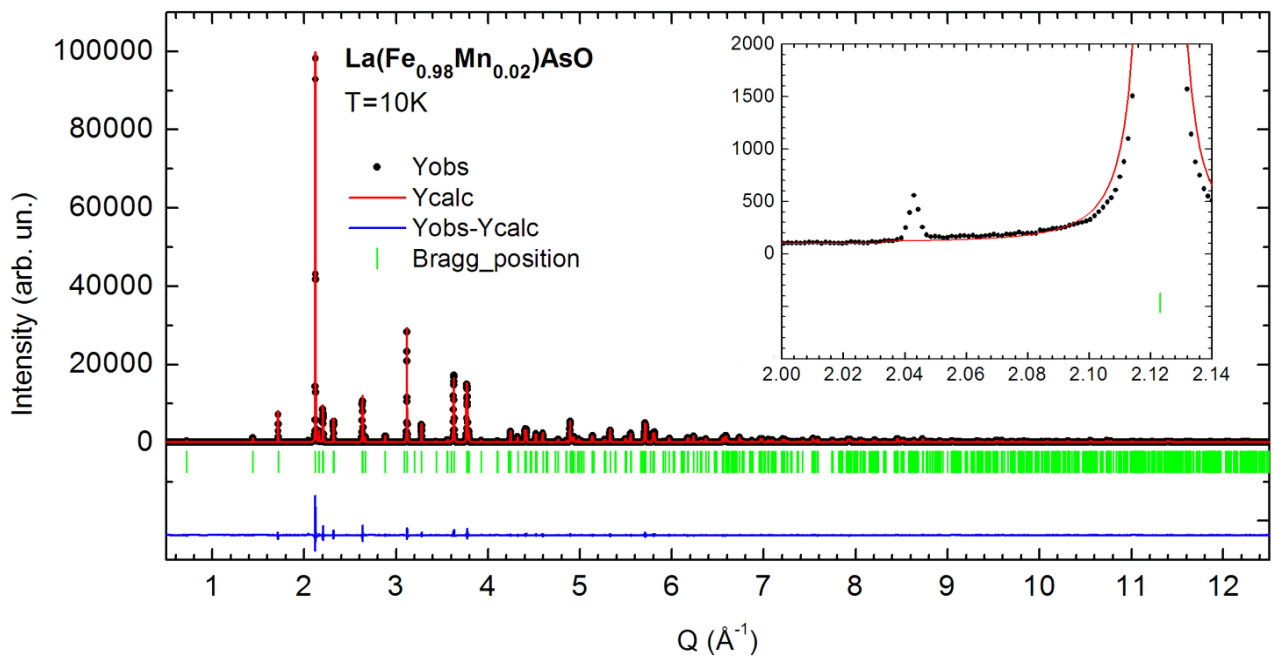

Figure 1: Rietveld refinement plot for La(Fe$_{0.98}$Mn$_{0.02}$)AsO (X-ray synchrotron data collected at 10 K); the black points represent the observed intensity data, the calculated pattern is superposed and drawn as a solid red line, the difference between the observed and calculated patterns is plotted in the lower field (blue line); the small vertical bars indicate the position of the allowed Bragg reflections for the orthorhombic structure. The inset shows a magnification of the plot, evidencing that the 1$^{st}$ order satellite peak of the modulated structure is not fitted by using a conventional *Cmme* structural model.

A closer inspection reveals that these additional reflections are not present at room temperature, whereas they arise after the establishment of the orthorhombic phase. These Bragg reflection peaks cannot be indexed by 3-integer indices, marking the occurrence of an incommensurate modulated structure. The intensities of these satellite reflections are about 10$^4$ times weaker than the main Bragg reflection; moreover, the intensity strongly decreases as the peak order increases and hence reflections up to 2$^{nd}$ order only can be detected (the one at 4.08 Å$^{-1}$). These satellite peaks are clearly detectable for $T \leq 75$ K, but a closer inspection at higher temperatures reveals that a significant satellite Bragg scattering can be detected also in the data collected at 105 K, but not at 110 K (Figure 2). These data suggest that the incommensurate structure sets in at a slightly, but significantly higher temperature than the spin density wave ordering ($T_{SDW} = 99.5$ K). The satellite peak intensities increase as the temperature decreases (Figure 3), indicating a progressive increase of the modulation amplitude, whereas the modulation vector remains constant in the whole inspected temperature range. From the width of the Bragg peak at Q ~ 2.04 Å$^{-1}$ and after correction for the instrument resolution, it can be estimated that the modulation has a correlation length $\xi \sim 75$ Å at 10 K. Remarkably, this is the first experimental evidence for the occurrence of an incommensurate structure in 1111-type Fe-based superconductors.

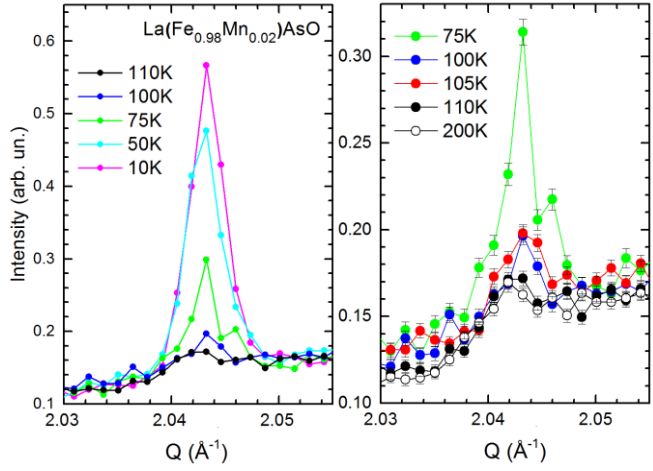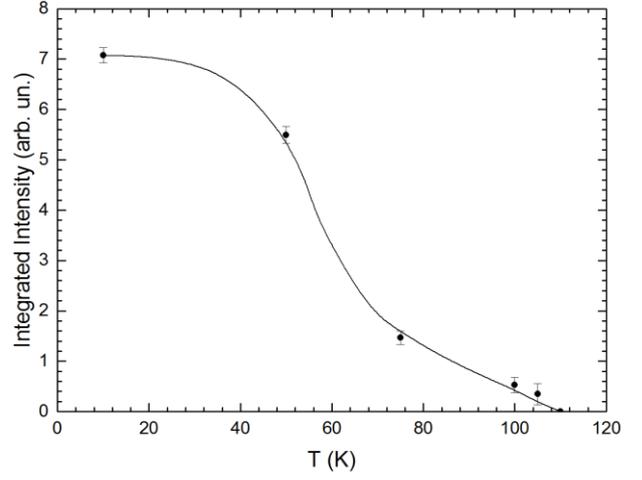

Figure 2: Thermal evolution of the 1st order satellite peak of La(Fe$_{0.98}$Mn$_{0.02}$)AsO.

Figure: 3 Thermal dependence of the satellite peak integrated intensity measured in La(Fe$_{0.98}$Mn$_{0.02}$)AsO (the line is a guide to the eye).

In the La(Fe$_{0.96}$Mn$_{0.04}$)AsO sample the structural transformation is further decreased down to ~ 90 K and only a 1st order satellite peak can be detected for $T \leq 30$ K at Q ~ 2.33 Å$^{-1}$ (Figure 4), indicating a different modulation of the crystal structure.

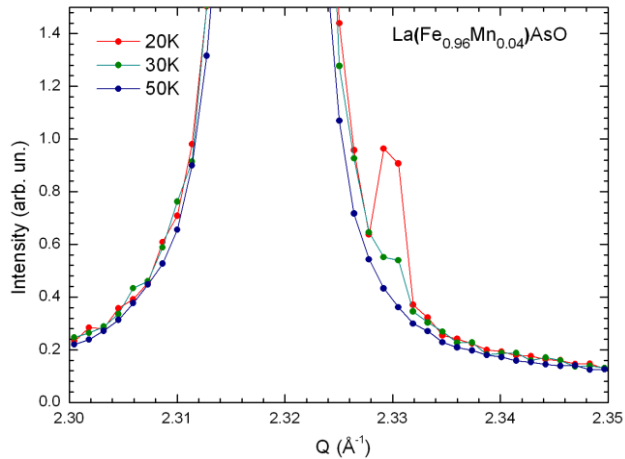

Figure 4: Thermal evolution of the 1st order satellite of La(Fe$_{0.96}$Mn$_{0.04}$)AsO.

The occurrence of a static charge density wave is also confirmed by HRTEM analysis carried out at low temperature on the same La(Fe$_{0.98}$Mn$_{0.02}$)AsO sample. As mentioned above, the modulated structure mostly develops below ~ 75 K, a temperature exceeding the instrumental lowest limit of our cryogenic TEM holder (94 K, nominal value). In any case at this temperature the 1st order tetragonal to orthorhombic structural transition is almost completed and some regions of the sample exhibit incipient formation of a static structural modulation along the crystal direction [100]. In particular, [ $0\bar{2}1$] HRTEM projections at cryo-temperature revealed {100} lamellae, 1-2 nm thick, within the undisturbed crystal lattice. These lamellae exhibited a more complex projected potential with respect to that of the host structure suggesting an incipient development of the aperiodic structure. Figure 5a shows a region of about 250 nm$^2$ of undisturbed orthorhombic crystal structure, with clear (200) lattice sets (interplanar spacing ~ 2.8 Å), where a structural modulation is evident on the right. Figure 5b displays a magnified view of modulated structure characterized by {100} lamellae, alternating approximately every 3 nm, and by {112} lattice fringes (interplanar spacing ~ 2.9 Å). The interface {100}-lamellae/{100}-orthorhombic structure exhibits a semi-coherent to coherent nature indicating a crystallographic continuity along {100} and {112} planes. In particular the {$h$00} periodicity of lamellae and {200} plane of orthorhombic structure is quite preserved.

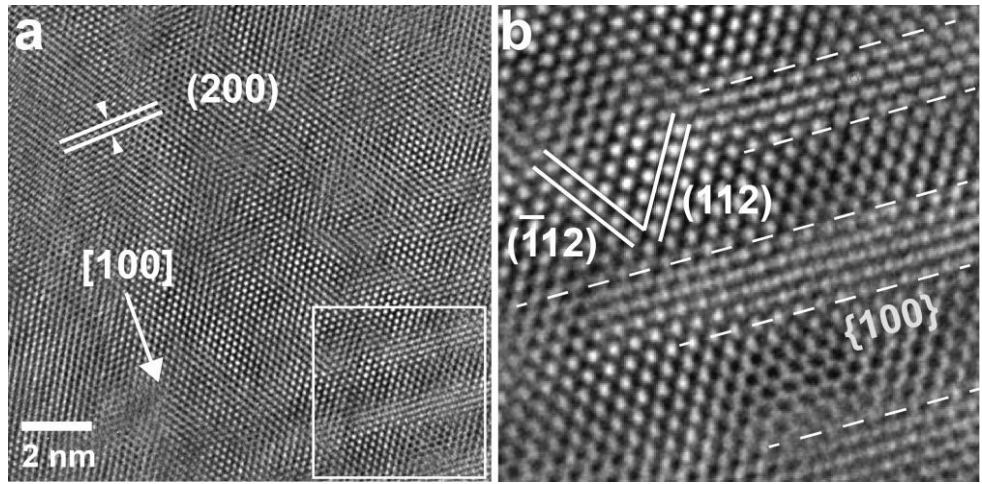

Figure 5: HRTEM image of La(Fe$_{0.98}$Mn$_{0.02}$)AsO phase at 94 K and observed along the [$0\bar{2}1$] zone axis (orthorhombic structure). a) HRTEM overview of undisturbed region evidencing {200} lattice sets with inter-planar distance of ~ 2.8 Å; in the box a small area with structural modulation. b) Magnified view of the marked area highlighting the {100} lamellae alternating along the [100] direction and the {112} lattice fringes.

The detection of a structural modulation in lightly Mn-substituted LaFeAsO samples represents a fundamental step towards the comprehension of the physics of Fe-SC. In this context it is worth noting that modulations can only be detected by X-ray or neutron scattering measurements when they are static (or otherwise when dynamic modulations are pinned by impurities); in any case satellite diffraction peaks can be also $10^{10}$ times weaker than the main Bragg reflections and hence their detection can be extremely challenging. Furthermore, static atomic modulations (involved in the CDW state) are usually characterized by a small magnitude, and hence hard to be directly detected, especially in powdered samples. These are probably the reasons why structural modulations, if present, eluded previous analyses and were never observed before in Fe-based superconductors.

Another possibility is suggested by the observation of a significant decrease of the satellite peak intensity with the increase of Mn-content; likely, the aperiodic modulation is best detected around an optimal (very light) concentration of impurities in the Fe-sublattice, an occurrence that is seldom investigated. As a rule, impurities interact and can pin a CDW, breaking the translational invariance, since their electronic properties substantially differ from the host atoms. In particular, an incommensurate CDW weakly interacts with the underlying lattice, but a significant variation of the lattice potential due to atomic impurities can effectively interact with the periodic charge modulation of the CDW; as a result the incommensurate CDW is distorted or pinned [16] and its translational motion is suppressed. Since that incommensurate modulation was never previously observed in other substituted Fe-based superconductors, it can be presumed that the pinning of the CDW is sensitive to the chemical nature of the impurities. On this basis, it can be conjectured that the CDW is likely pinned by the characteristic nature of the substituting $Mn^{2+}$ ion (a $3d^5$ species) in our samples; in fact, the strong Hund's coupling aligns all the five $d$-electrons, yielding a half-filled $d$-shell and a highly stable electronic configuration that prevents spin and valence fluctuations affecting the surrounding Fe-sublattice. Hence, these ions are not involved in collective dynamic electronic states (charge or spin density wave states) characterizing the Fe sub-lattice, but conversely can act as very effective pinning centres, producing static correlations. Resistivity measurements performed on the same La(Fe$_{1-x}$Mn$_x$)AsO sample series ($x$ = 0.00, 0.01, 0.02 and 0.04; see supplementary information) exhibit an abrupt upturn in the temperature region where the structural modulation is clearly detectable. This evidence is consistent with an active role of Mn impurities as pinning centres for the CDW.

Our results indicate that the Fermi surface nesting can induce a CDW state in these materials, responsible for the structural transition. Remarkably the tetragonal to orthorhombic transformation induces a rotation by 45° of the unit cell; hence the in-plane $\mathbf{k}_{nesting} = (\pi,\pi)$ calculated for the high temperature tetragonal phase is parallel to the in-plane $\mathbf{k}_{modulation} = (\delta,0)$ detected for the low-temperature orthorhombic phase by HRTEM analysis. This is the exact situation expected for a CDW state.

As pointed out by Kivelson and Emery [17] SDW order always implies CDW order, even though the latter is much harder to detect and can be only inferred in most cases by the occurrence of the former. Remarkably, when CDW takes place at a higher temperature than SDW (or when SDW is completely absent), the density wave transition is charge-driven; conversely, the transition is spin-driven when both CDW and SDW develop simultaneously [17]. As afore mentioned, in La(Fe$_{0.98}$Mn$_{0.02}$)AsO the structural modulation appears at a slightly higher temperature than the magnetic ordering, indicating that the two transitions are decoupled in temperature. We underline the fact that further experiments are required in order to collect more data points around the critical temperature and gain more firmly conclusions. Nonetheless, on the basis of the present data, it can be argued that the density wave ordering is charge-driven and consequently the structural transition as well; Table 1 summarize the critical transition temperatures observed in La(Fe$_{0.98}$Mn$_{0.02}$)AsO.

Table 1: Critical transition temperatures observed in La(Fe$_{0.98}$Mn$_{0.02}$)AsO as determined by synchrotron X-ray (SXRPD) and neutron powder diffraction data (NPD).

|  |  |
|---|---|
| $T_s$ | 115±2.5 K (SXRPD) |
|  | 113.5±0.5 K (NPD) |
| $T_{CDW}$ (K) | 107.5±2.5 K (SXRPD) |
| $T_{SDW}$ (K) | 99.5±0.5 K (NPD) |

These findings probably reconcile the apparently contradictory behaviour observed in FeSe as well as in many underdoped 122- and 1111-type compounds, characterized by an orthorhombic structure at low temperature and a fully superconductive state, where magnetism is completely suppressed. It could be envisioned that in all these materials a CDW state occurs, yielding the low temperature structural distortion. In any event, more focused investigations are needed to corroborate this scenario.

Summarizing, our results disclose a new view for the phenomenology and phase diagrams of Fe based superconductors. Our observation of a static CDW in La(Fe$_{1-x}$Mn$_x$)AsO reveals the possibility of an intrinsic and widespread tendency towards charge ordering in Fe-based superconductors. The possible role of CDW fluctuations for the mechanism of the nematic and orthorhombic phase formation as well as their interplay with superconductivity represent new relevant subjects for both experimental and theoretical researches. The huge suppression of superconductivity characterizing both La(Fe$_{1-x}$Mn$_x$)As(O$_{0.89}$F$_{0.11}$) and (La$_{2-x}$Ba$_x$)CuO$_4$ systems (and some related systems as well) coupled with the establishment of static magnetism in both kinds of materials, suggests that alike phenomena are at play in both Cu- and Fe-based superconductors. Remarkably, charge density wave and superconductive states are often adjacent and competing in the phase diagrams of several unconventional superconducting materials, such as Cu-based superconductors, layered transition metal chalcogenides and A15 compounds [18].

**Acknowledgements**

A. M acknowledges Andy Fitch for his kind support and experimental assistance during the data collection at the ID22 beamline of ESRF (proposal HC-1468).
9

**References**

[1] Fernandes, R. M., Chubukov, A. V., Schmalian, J., Nature Physics **10**, 97 (2014)

[2] Singh, D. J., Du, M.-H., Phys. Rev. Lett. **100**, 237003 (2008)

[3] Mazin, I. I., Singh, D. J., Johannes, M. D., Du, M. H., Phys. Rev. Lett. **101**, 057003 (2008)

[4] Dong, J., Zhang, H. J., Xu, G., Li, Z., Li, G., Hu, W. Z., Wu, D., Chen, G. F., Dai, X., Luo, J. L., Fang, Z., Wang, N. L., **83**, 27006 (2008)

[5] de la Cruz, C., Huang, Q., Lynn, J. W., Li, J., Ratcliff II, W., Zarestky, J. L., Mook, H. A., Chen, G. F., Luo, J. L., Wang, N. L., Dai, P., Nature **453**, 899 (2008)

[6] Fang, C., Yao, H., Tsai, W-F., Hu, J., Kivelson, S. A., Phys. Rev. B **77**, 224509 (2008)

[7] Xu, C., Muller, M., Sachdev, S., Phys. Rev. B **78**, 020501(R) (2008)

[8] A. Martinelli, J. Phys.: Condens. Matter 25 (2013) 125703

[9] Lee, S.-H., Xu, G., Ku, W., Wen, J.S., Lee, C.C., Katayama, N., Xu, Z.J., Ji, S., Lin, Z.W., Gu, G.D., Yang, H.-B., Johnson, P.D., Pan, Z.-H., Valla, T., Fujita, M., Sato, T.J., Chang, S., Yamada, K., Tranquada, J.M., Phys. Rev. B **81**, 220502(R) (2010)

[10] Hammerath, F., Bonfà, P., Sanna, S., Prando, G., De Renzi, R., Kobayashi, Y., Sato, M., Carretta, P., Phys. Rev. B **89**, 134503 (2014)

[11] Sato, M., Kobayashi, Y., Lee, S. C., Takahashi, H., Satomi, E., Miura, Y., J. Phys. Soc. Jpn. **79**, 014710 (2010)

[12] Tranquada, J. M., Sternlieb, B. J., Axe, J. D., Nakamura, Y., Uchida, S., Nature **375**, 561 (1995)

[13] Tranquada, J. M., Axe, J. D., Ichikawa, N., Nakamura, Y., Uchida, S., Nachumi, B., Phys. Rev. B **54**, 7489 (1996)

[14] Tranquada, J. M., Axe, J. D., Ichikawa, N., Moodenbaugh, A. R., Nakamura, Y., Uchida, S., Coexistence of, and Competition between, Phys. Rev. Lett. **78**, 338 (1997)

[15] Rodríguez–Carvajal, J., Physica B **192**, 55 (1993)

[16] Dai, H., Liu, J., Lieber, C. M., Elucidating Complex Charge Density Wave Structures in Low-Dimensional Materials by Scanning Tunneling Microscopy, in Advances in the Crystallographic and Microstructural Analysis of Charge Density Wave Modulated Crystals, Boswell, F.W., Bennett, J. Craig (Eds.), Kluwer Academic Publishers (1999)

[17] Kivelson, S. A., Emery, V. J., Stripe Liquid, Crystals, and Glass Phases of Doped Antiferromagnets, in Stripes and Related Phenomena, edited by Bianconi and Saini, Kluwer Academic/Plenum Publishers, New York (2000)

[18] Gabovich, A. M., Voitenko, A. I., Annett, J. F., Ausloos, M., Supercond. Sci. Technol., **14**, R1 (2001)

SUPPLEMENTARY INFORMATION

**Experimental evidences for static charge density waves in iron oxy-pnictides**

*S.1: Synchrotron powder diffraction*

The structural properties of La(Fe$_{1-x}$Mn$_x$)AsO samples ($x = 0.02$ and 0.04) were investigated by means of high-resolution synchrotron X-ray powder diffraction analysis between 290 K and 10 K; X-ray scattering data were collected at the ID22 high-resolution powder diffraction beamline of the European Synchrotron Radiation Facility (ESRF) in Grenoble, France. Structural data at 290 K and 10 K obtained after Rietveld refinement are listed in Tables S1 and S2.

Table S1: Structural data at 290 K obtained for La(Fe$_{1-x}$Mn$_x$)AsO samples after Rietveld refinement in the *P*4/*nmm* space group (synchrotron X-ray diffraction data).

|  |  | La(Fe$_{0.98}$Mn$_{0.02}$)AsO | | | La(Fe$_{0.96}$Mn$_{0.04}$)AsO | | |
|---|---|---|---|---|---|---|---|
| Cell parameters | *a* (Å) | 4.0372(1) | | | 4.0391(1) | | |
|  | *c* (Å) | 8.7505(1) | | | 8.7640(1) | | |
| Atomic site |  | *x* | *y* | *z* | *x* | *y* | *z* |
| La | 2*c* | ¼ | ¼ | 0.1411(1) | ¼ | ¼ | 0.1409(1) |
| (Fe,Mn) | 2*a* | ¾ | ¼ | ½ | ¾ | ¼ | ½ |
| As | 2*c* | ¼ | ¼ | 0.6518(1) | ¼ | ¼ | 0.6521(1) |
| O | 2*b* | ¾ | ¼ | 0 | ¾ | ¼ | 0 |
| $R_F$ (%) |  | 4.50 | | | 3.38 | | |
| $R_{Bragg}$ (%) |  | 8.75 | | | 2.97 | | |

Table S2: Structural data at 10 K obtained for La(Fe$_{1-x}$Mn$_x$)AsO samples after Rietveld refinement in the *Cmme* space group (synchrotron X-ray diffraction data).

|  |  | La(Fe$_{0.98}$Mn$_{0.02}$)AsO |  |  | La(Fe$_{0.96}$Mn$_{0.04}$)AsO |  |  |
|---|---|---|---|---|---|---|---|
| Cell parameters | $a$ (Å) | 5.7066(1) |  |  | 5.7066(1) |  |  |
|  | $b$ (Å) | 5.6911(1) |  |  | 5.7066(1) |  |  |
|  | $c$ (Å) | 8.7206(1) |  |  | 8.7310(1) |  |  |
| Atomic site |  | $x$ | $y$ | $z$ | $x$ | $y$ | $z$ |
| La | 4$g$ | 0 | ¼ | 0.1417(1) | 0 | ¼ | 0.1415(1) |
| (Fe,Mn) | 4$b$ | ¼ | 0 | ½ | ¼ | 0 | ½ |
| As | 4$g$ | 0 | ¼ | 0.6511(1) | 0 | ¼ | 0.6514(1) |
| O | 4$a$ | ¼ | 0 | 0 | ¼ | 0 | 0 |
| $R_F$ (%) |  | 5.43 |  |  | 4.97 |  |  |
| $R_{Bragg}$ (%) |  | 4.73 |  |  | 4.90 |  |  |

Generally speaking, the variation of the lattice parameters of incommensurately modulated structures may induce microstrain-like anisotropic line broadening in powder diffractions patterns[18]. In this context, Figure S1 shows the superposition of the Williamson-Hall plots obtained after Rietveld refinements of the data collected at 10 K and 290 K ($x = 0.02$); for a sake of clarity, data at 10 K are reported according to the tetragonal setting. It is evident that low temperature data are affected by a micro-strain like contribution to the line broadening. From these data it can be concluded that microstrain increases by almost 30% along [00$l$] and <$h$00>, but the gain exceeds 50% along <$hh$0>, which represents the direction of the Fermi surface nesting wave vector.

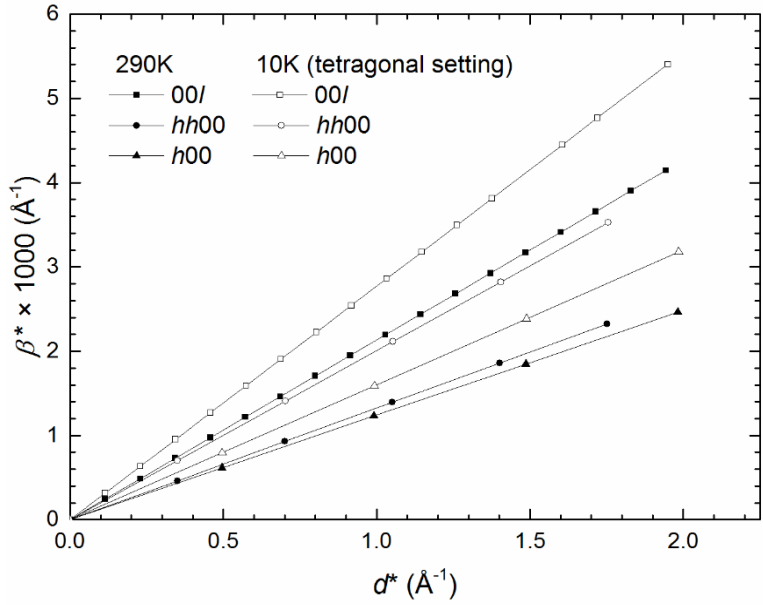

Figure S1: Superposition of the Williamson-Hall plots obtained after Rietveld refinements of the data collected at 10 K and 290 K for the La(Fe$_{0.98}$Mn$_{0.02}$)AsO sample.

This behaviour complies with the case where only the parameters describing the average lattice vary in the different crystallites, but not the modulation vector$^{\text{Errore. Il segnalibro non è definito.}}$. This is confirmed by the fact that the position of the satellite reflection remains constant with temperature.

*S.2: Determination of the magnetic transition temperature*

Neutron diffraction analysis is the best and direct probe of a magnetically ordered state. Preliminary data were collected for the La(Fe$_{0.98}$Mn$_{0.02}$)AsO sample using the high-intensity medium-resolution neutron powder D20 diffractometer of the Institute Laue-Langevin ($\lambda$ = 2.41 Å). Remarkably, by these neutron diffraction data, also the structural transition temperature can be determined to $T_s$ = 113.5±0.5 K, in optimal agreement with the result obtained by synchrotron X-ray powder diffraction data. Figure S2 shows the thermal evolution of the magnetic moment as obtained after Rietveld refinement using the high-intensity neutron powder diffraction data; a $T_{\text{SDW}}$ = 99.5±0.5 K results by fitting these data using a Landau mean field model $m(T) = m_0(1 - T/T_m)^\beta$, where $m_0$ is the magnetic moment extrapolated at $T$ = 0 K and $\beta$ is the critical exponent, which is ½ in the case of pure 2$^{\text{nd}}$ order transitions. This result complies with the magnetization measurement; Figure S.3 shows the magnetization curve and the corresponding *erf* fit; the inflection point is located at 104.5±0.4 K, a few degrees above the long range AF order probed by neutrons.

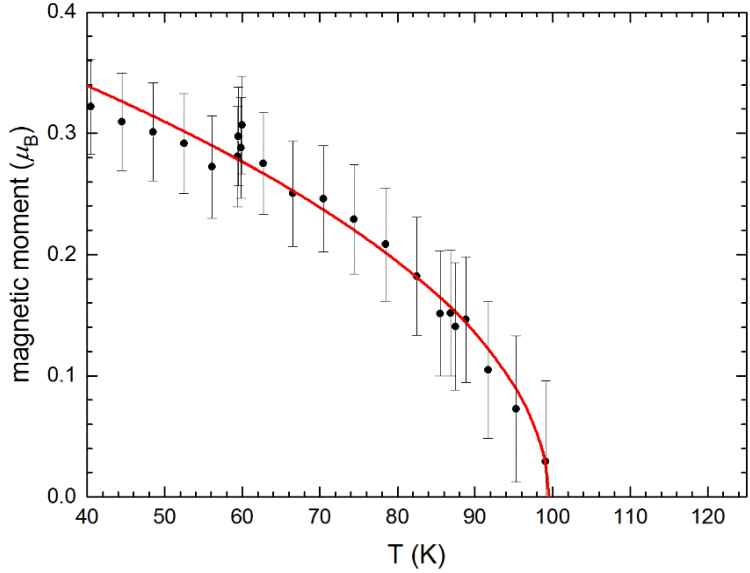

Figure S2: Thermal evolution of the magnetic moment in the La(Fe$_{0.98}$Mn$_{0.02}$)AsO sample as obtained after Rietveld refinement of high-intensity low-resolution neutron powder diffraction data; the data are fitted according to a Landau mean field theory valid for a 2$^{nd}$ order transition (red line).

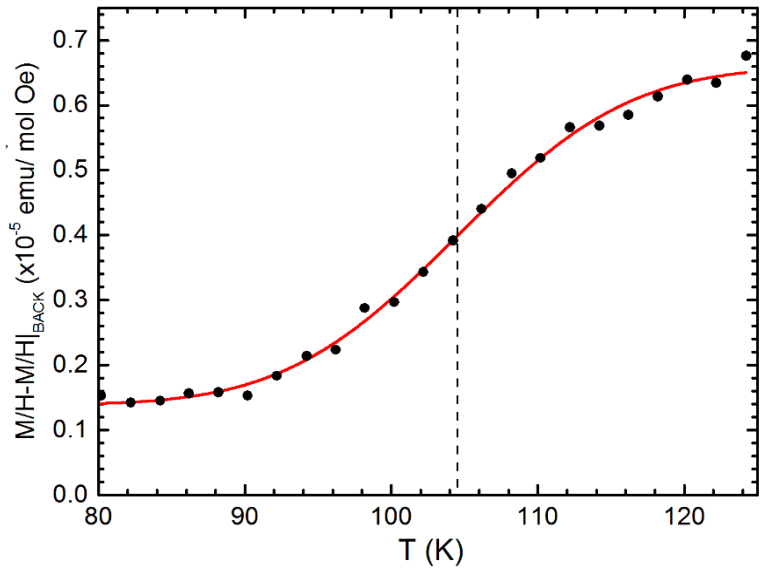

Figure S3: Temperature dependence of M/H measured under an applied field of 3 T in the magnetic transition region after subtracting a Curie–Weiss-like 'background' (M/H|$_{BACK}$);[18] the red line represents the *erf*-fit whose inflection point is at $T \sim 104.5$ K (dashed line).

*S.3: dc resistivity*

Figure S4 shows the electrical resistivity (normalized to its value at T = 300 K) measured as a function of the temperature. The increase of the Mn content causes a suppression of structural/magnetic ordering temperature with a progressive smoothing of the typical step-like feature that characterizes the $x = 0$ composition around $T = 150$ K (blue curve in Figure S4). Moreover, the substitution of even a small amount of Mn substantially changes the nature of conduction. In fact in the $x = 0.01$ sample, the resistivity which decreases below the magnetic/structural transition, undergoes an abrupt upturn below 50 K. In the heavier substituted samples ($x = 0.02$ and 0.04) the low temperature upturn appears superimposed to a semiconducting-like behaviour starting at room temperature.

The inset of Figure S4 shows the evolution of the Hall coefficient curves, R$_H$ vs T, as a function of the Mn content. The increase of the Mn content causes a suppression of the structural/magnetic ordering temperature and a progressive diminution of the absolute value, which can be ascribed to a compensation effect characteristic of a multiband system.

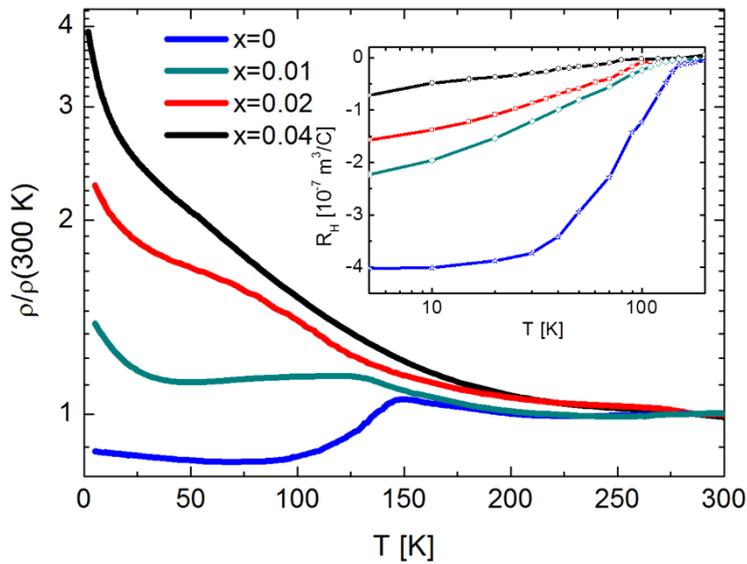

Figure S4: Normalized resistivity $\rho/\rho(300\text{ K})$ vs T measurement of La(Fe$_{1-x}$Mn$_x$)AsO ($x$ =0, 0.01, 0.02 and 0.04) series; inset: Hall coefficient R$_H$ vs T measurement of La(Fe$_{1-x}$Mn$_x$)AsO ($x$ =0, 0.01, 0.02 and 0.04) series.

**References**

A. Leineweber, V. Petricek, Microstrain-like diffraction-line broadening as exhibited by incommensurate phases in powder diffraction patterns, J. Appl. Cryst. **40**, 1027 (2007)

[18] M. A. McGuire et al., New J. Phys. 11, 025011 (2009).